# Polarisation-Sensitive THz Detectors

M.B. Johnston, E. Castro-Camus, J. Lloyd-Hughes, M.D. Fraser[1], H.H. Tan[1] and C. Jagadish[1],

Department of Physics, University of Oxford, Clarendon Laboratory, Parks Road, Oxford OX1 3PU, United Kingdom,
[1] Department of Electronic Materials Engineering, Research School of Physical Sciences and Engineering, Institute of Advanced Studies, Australian National University, Canberra ACT 0200.
e-mail: M.Johnston@physics.ox.ac.uk

**Abstract**

We have developed a detector of coherent terahertz (THz) radiation that can recover the full polarisation state of a THz transient. The device is three-contact photoconductive receiver, which is capable of recording two time-varying electric field components of a THz pulse simultaneously. Our receiver was fabricated on $Fe^+$ implanted InP and showed a cross-polarised extinction ratio greater than 100:1. The detector will be useful for spectroscopy of birefringent and optically active materials.

**Introduction**

Terahertz time domain spectroscopy (THz-TDS) is a versatile technique for spectroscopy in the far, mid and recently near-infrared regions of the spectrum [1]. What sets THz-TDS apart from other forms of spectroscopy is the ability to measure directly the time-varying electric field of an electromagnetic wavepacket. Thus the form of the wavepacket is experimentally recorded, without the need to measure multiple orders of its autocorrelation function. Furthermore the complete dielectric function of a material under investigation may be extracted straightforwardly [2]. In contrast most conventional forms of optical spectroscopy are reliant on detectors that record intensity rather the electric field, thus information about phase of the electric field is not directly measured.

While THz-TDS has already becoming a valuable tool for pure and applied research [3, 4] the technique has not been fully exploited for spectroscopy of birefringent and optically active materials. Recent free electron laser studies have shown strong THz circular dichroism signal from biologically significant molecules [4]. Thus, there is great potential for using THz-TDS in structure-and-function studies of chiral biological molecules. Such TDS measurements would require measurement of the full polarisation state of a THz pulse.

To date THz-TDS experiments have only been able to resolve one electric field component of a transient at a time. The ability to obtain the direction and magnitude of a THz pulse's electric field vector as a function of time in a single measurement would allow us to completely characterise a THz pulse. Such a measurement would be made possible by simultaneously measuring two orthogonal components of the THz electric field as a function of time. We have developed a detector that is capable of achieving this.

**Detector Design and Fabrication**

In order to measure two orthogonal components of THz electric field simultaneously we designed a three-contact photoconductive receiver [5]. Photoconductive receivers and electro-optic sampling are the two main detection methods used in present THz-TDS spectrometers [4].

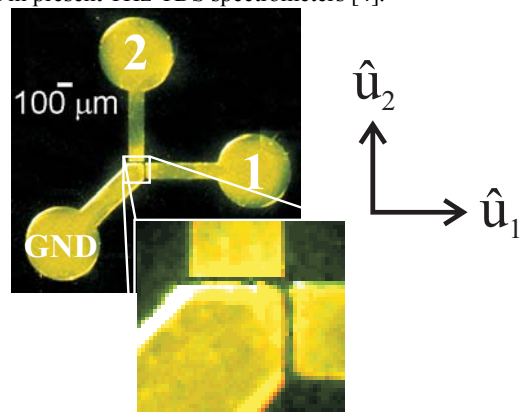

**Fig. 1**: Photograph of polarisation sensitive photoconductive receiver. The circular regions are pads to facilitate ultrasonic wire-bonding. The inset shows an enlarged image of the two orthogonal gaps. The common terminal is labelled GND and the electrode for measuring the component of THz electric field in the $u_{1(2)}$–direction is labelled 1 (2).

Conventional photoconductive receivers consist of two contacts usually incorporating some form of antenna structure. The most commonly used receivers include a tuned strip-line antenna or broadband bow-tie antenna [4]. However, these devices are only capable of recording one polarisation component of THz electric field at a time.

The three-contact photoconductive receiver was fabricated on $Fe^+$-ion implanted semi-insulating InP (100) substrates. The $Fe^+$ ions were implanted at two different energies (2.0MeV and 0.8MeV) and doses (1.0 x $10^{13}$ cm$^2$ and 2.5 x $10^{12}$ cm$^2$ respectively) so as to give an approximately even vacancy distribution to a depth of 1 μm from the surface. The samples were subsequently annealed at 500°C for 30 minutes under a $PH_3$ atmosphere. The three electrodes were defined using standard photolithography and lift-off techniques. The Cr/Au contacts were deposited to a thickness of 20/250nm using a thermal evaporator.

An optical micrograph of the device used in this study is shown in Fig. 1. The device consists of two orthogonal 16-micron gap electrodes. It was mounted into a chip package and the contacts were bonded with gold wires using an ultrasonic bonder. A gold-coated lid, with a 1mm diameter hole drilled though it, was used to enclosed the device within the package. The hole was positioned directly about the gap region of the device.

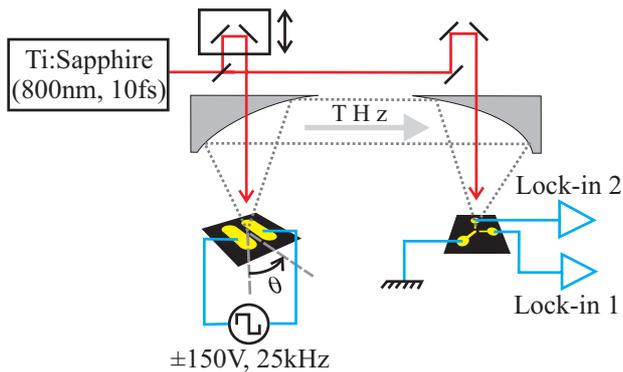

**Fig. 2:** Schematic diagram of THz-TDS system used to characterize the three-contact photoconductive receiver.

### Experiment

In order to characterise the performance of the three-contact receiver we tested its response to linearly polarised THz radiation at a series of angles of polarisation. A schematic diagram of the THz-TDS system used for this demonstration is shown in Fig 2. Details of the experimental set-up may be found in Ref [5]. Briefly, a 400 μm gap photoconductive switch [6] was used as a source of linearly polarised THz radiation. A 25kHz, 300V peak-to-peak square wave was applied across the emitter, which was excited using a 4nJ, 10fs pulses from a mode locked Ti:sapphire laser (75MHz repetition rate). The linearly polarized THz source was placed in a rotation stage, thereby allowing the angle of polarization to be set arbitrarily. The THz pulse was collected and then focused onto the test receiver using two off-axis parabolic mirrors. Two separate lock-in amplifiers were used to record the current flowing between earth and contacts 1 and 2.

### Results

Fig. 3 shows the THz electric field measured with the three-contact receiver at three different emitter polarisation angles (0, 45 and 90 degrees). The electric field signal is the derivative of the gap current in device (see Ref [5] for details). As seen in the figure, the receiver accurately records the polarisation change of the source. In order to assess the performance of the device we measured the cross-polarised extinction ratio of the system. The ratios were measured to be 108:1 and 128:1 for a horizontally and vertically aligned emitter respectively. However these values probably represent a lower limit to the device performance since this experiment is susceptible to slight misalignment of the off axis parabolic mirrors and the presence of a small quadrupole component in the source radiation.

The signal-to-noise ratio in these experiments was 175:1, which was similar to that measured for a conventional bow-tie detector fabricated on the same substrate. This indicates that the signal-to-noise ratio in our receivers is limited by the substrate and/or fabrication processes rather than by an intrinsic feature of the device design.

### Conclusions

We have demonstrated a photoconductive receiver that can measure the full polarization state of a THz pulse. This device will have immediate application for THz-TDS experiments on birefringent and optically active samples. We expect the signal-to-noise performance of the device to improve significantly with the used of optimised implanted InP or low temperature grown GaAs substrates.

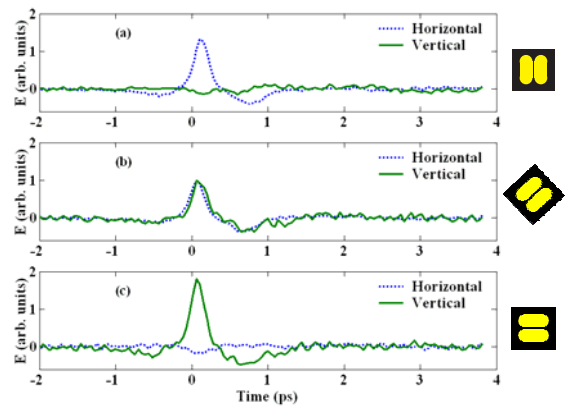

**Fig. 3:** THz electric field as a function of time recorded using a three-contact photoconductive receiver. The horizontal and vertical components were recorded simultaneously on two separate lock-in amplifiers. The schematic diagrams to the right of the panel indicates the orientation, and hence polarisation of the photoconductive emitter.

### Acknowledgements

The authors would like to thank the EPSRC (UK) and the Royal Society for financial support of this work, ECC wishes to thank CONACyT (Mexico) for a scholarship. Australian authors would like to acknowledge the financial support of the Australian Research Council.